\documentclass[aps,pra,letterpaper,superscriptaddress,showpacs,amsmath,floats,twocolumn]{revtex4-1}

\usepackage{graphicx}
\usepackage{amssymb,amsfonts,amsmath}
\usepackage{physics}
\usepackage{siunitx}
\usepackage{braket}
\usepackage{bm}
\usepackage[utf8]{inputenc}
\usepackage[T1]{fontenc}
\usepackage{lmodern}
\usepackage{mathtools}
\usepackage{alltt}
\usepackage{mathrsfs}

\begin{document}

\title{Ultra-high Compton Frequency, Parity Independent, Mesoscopic Schr\"odinger Cat Atom Interferometer with Heisenberg Limited Sensitivity}

\author{Resham Sarkar}
\affiliation{Department of Physics and Astronomy, Northwestern University, 2145 Sheridan Road, Evanston, IL 60208, USA}

\author{Renpeng Fang}
\affiliation{Department of Physics and Astronomy, Northwestern University, 2145 Sheridan Road, Evanston, IL 60208, USA}

\author{Selim M. Shahriar}
\affiliation{Department of Physics and Astronomy, Northwestern University, 2145 Sheridan Road, Evanston, IL 60208, USA}
\affiliation{Department of EECS, Northwestern University, 2145 Sheridan Road, Evanston, IL 60208, USA}
\email{shahriar@northwestern.edu}

\date{\today}

\begin{abstract}
We present a protocol for an atomic interferometer that reaches the Heisenberg Limit (HL), within a factor of $\sim$ $\sqrt{2}$, via collective state detection and critical tuning of one-axis twist spin squeezing. It generates a Schr\"odinger cat (SC) state, as a superposition of two extremal collective states. When this SC interferometer is used as a gyroscope, the interference occurs at an ultrahigh Compton frequency, corresponding to a mesoscopic single object with a mass of $Nm$, where $N$ is the number of particles in the ensemble, and $m$ is the mass of each particle. For $^{87}$Rb atoms, with $N=10^{6}$, for example, the intereference would occur at a Compton frequency of $\sim$ $2 \times 10^{31}$ Hz. Under this scheme, the signal is found to depend critically on the parity of $N$. We present two variants of the protocol. Under Protocol A, the fringes are narrowed by a factor of $N$ for one parity, while for the other parity the signal is zero. Under Protocol B, the fringes are narrowed by a factor of $N$ for one parity, and by a factor of $\sqrt{N}$ for the other parity. Both protocols can be modified in a manner that reverses the behavior of the signals for the two parities. Over repeated measurements under which the probability of being even or odd is equal, the averaged sensitivity is smaller than the HL by a factor of $\sim$ $\sqrt{2}$ for both versions of the protocol. We describe an experimental scheme for realizing such an atomic interferometer, and discuss potential limitations due to experimental constraints imposed by the current state of the art, for both collective state detection and one-axis-twist squeezing. We show that when the SC interferometer is configured as an accelerometer, the effective two-photon wave vector is enhanced by a factor of $N$, leading to the same degree of enhancement in sensitivity.  We also show that such a mesoscopic single object can be used to increase the effective base frequency of an atomic clock by a factor of $N$, with a sensitivity that is equivalent to the HL, within a factor of $\sim$ $\sqrt{2}$.  
\end{abstract}


\maketitle


\section{Introduction}
\label{Introduction}
The phase sensitivity of an atomic interferometer (AI), when used as a gyroscope, depends on the Compton frequency, $\omega_c = mc^2/\hbar$ of the individual particles interfering at non-relativistic velocities, where $m$ is the mass of the particle, and $c$ is the velocity of light in vacuum~\cite{Borde, Chu, Riehle, RefComment}. Matter wave interferometry with large molecules have successfully demonstrated the superposition of quantum states with large mass~\cite{Eibenberger}. However, these interferometers, based on the Talbot effect, are not suited for rotation sensing, owing to constraints in fabricating gratings of small enough spacing, and associated effects of van der Waals interaction. An alternative approach is to make a large number ($N$) of particles, each with a mass $m$, behave as a single object with a mass of $M\equiv Nm$, and thus a Compton frequency of $Mc^2/\hbar$. For a million $^{87}$Rb atoms, for example, this frequency is $\sim$ $2 \times 10^{31}$ Hz. In this paper, we describe a protocol that enables the realization of an atomic interferometer where two distinct quantum states of such a mesoscopic single object, each with this Compton frequency, are spatially separated and then recombined, leading to fringes that are a factor of $N$ narrower than what is achieved with a conventional atomic interferometer. We show that the net metrological sensitivity of this interferometer is equivalent to the Heisenberg limited (HL) sensitivity, within a factor of $\sqrt{2}$, of a conventional atomic interferometer. Aside from application to metrology, such a mesoscopic Schr\"odinger cat (SC)~\cite{Schroedinger} interferometer may serve as a test-bed for the effect of gravitational interaction on macroscopic decoherence and quantum state reduction~\cite{Diosi, Penrose1, Penrose2, Penrose3, Pikovski}. It also opens up a new regime for exploring performance of matter-wave clocks~\cite{Lan} in a regime with a much higher Compton frequency. 

When an AI is configured as an accelerometer, its sensitivity does not depend on the Compton frequency.  For a conventional Raman atomic interferometer (CRAIN), for example, the phase shift is proportional to the effective, two-photon wavevector, $k_{eff}$, given by the sum of the wavevectors of the fields used in producing the Raman excitation.  We show that, for the mesoscopic SC interferometer proposed here, the corresponding wavevector is given by $Nk_{eff}$, so that the fringes in this case are also narrowed by a factor of $N$.  As such, the net metrological sensitivity of the SC interferometer, when used as an accelerometer, is also equivalent to the HL sensitivity, within a factor of $\sqrt{2}$, of a conventional atom interferometric accelerometer.  We also show that such a mesoscopic SC state can be used to increase the effective base frequency of an atomic clock by a factor of $N$, with a sensitivity that is equivalent to the HL, within a factor of $\sqrt{2}$, of a conventional atomic clock. 

Recently, we presented a Collective State Atomic Interferometer (COSAIN)~\cite{COSAIN}, where we showed that the effect of large Compton frequency (when it is configure as a gyroscope) can be observed indirectly by detecting one of the collective states. These states, $\{\ket{E_0}, \ket{E_1}, \ldots, \ket{E_N}\}$, commonly referred to as the Dicke collective states, arise as a result of interaction of an ensemble of identical independent atoms with a semiclassical field~\cite{CollectiveDescription, Dicke, Arecchi}. The interferences between all of the collective states lead to a reduction in signal linewidth by a factor of $\sqrt{N}$ as compared to a conventional Raman atomic interferometer (CRAIN). However, this reduction by a factor of $\sqrt{N}$ in linewidth is countered by a corresponding reduction in the effective signal to noise ratio (SNR) since the system now behaves as a single particle. Therefore, the metrological sensitivity of a COSAIN is, under ideal conditions, the same as that of a CRAIN. A direct transition $\ket{E_0}\leftrightarrow\ket{E_N}$, bypassing all the intermediate collective states, would result in a signal of linewidth narrowed by a factor of $N$, thus yielding HL phase sensitivity despite the reduced SNR. However, there is no electric dipole coupling between $\ket{E_0}$ and $\ket{E_N}$ for non-interacting atoms, thus excluding the possibility to achieve this goal with conventional excitation.

Here, we propose a new protocol that employs squeezing and a rotation, followed by another rotation and unsqueezing~\cite{Yurke, Toscano, Goldstein} in a COSAIN to attain the HL phase sensitivity, within a factor of $\sqrt{2}$. Explicitly, we apply one axis twist (OAT) spin squeezing~\cite{Kitagawa,Schleier,Leroux1,Leroux2,Gil,Guo,Sondberg} around the $\vu{z}$ axis (defined as the spin-up direction) immediately following the first $\pi/2$-pulse in a CRAIN, which aligns the mean spin vector along the $\vu{y}$ axis. Prior to the application of the squeezing interaction, the population of the collective states follow a binomial distribution, corresponding to the Coherent Spin State (CSS)~\cite{Arecchi}. As the strength of squeezing is increased, the distribution begins to flatten out, eventually generating a Schr\"odinger cat state corresponding to an equal superposition of $\ket{E_0}$ and $\ket{E_N}$~\cite{Molmer} when the OAT squeezing is followed by a $\pi/2$ rotation around the $\vu{x}$ axis. The usual dark-$\pi$-dark sequence follows, at the end of which we apply a corrective rotation by $\pi/2$ (rather than $-\pi/2$, due to the state inversion caused by the $\pi$-pulse) around the $\vu{x}$ axis, and then apply a corrective reverse-OAT squeezing interaction about the $\vu{z}$ axis. Finally, the last $\pi/2$ pulse effectuates interference between the collective states, and the signal is detected by measuring the population of one of the collective states. Since the process makes use of a superposition of two mesoscopic quantum states, we name this a Schr\"odinger Cat Atomic Interferometer (SCAIN).

In recent years, much theoretical and experimental work have been carried out to improve the precision of atomic sensors using quantum non-demolition (QND) measurements or spin-squeezing, both of which generate entanglement among the atoms. For example, a reduction in variance by a factor of $5.6$ dB was observed in reference~\cite{Leroux1} using $5 \times 10^{4}$ atoms, using cavity assisted OAT spin squeezing. In reference ~\cite{Monika}, the maximum reduction in variance observed was $8.8$ dB, also using $5 \times 10^{4}$ atoms, but employing QND measurement.  In reference~\cite{Thompson}, a suppression of variance by a factor of $10.5$ dB was achieved for $4.8 \times10^{5}$ atoms, using QND measurement. In reference~\cite{Hosten}, a reduction in variance by a factor of $20.1$ dB was observed for $5 \times 10^{5}$ atoms, using a combination of OAT spin squeezing followed by a QND measurement. While these results are impressive and encouraging, it should be noted that the degree of improvement achieved is far below the HL, under which the variance is reduced by a factor of $N$ compared to the standard quantum limit (SQL); for $N=5 \times 10^{5}$, for example, this would correspond to a suppression of variance by a factor of $57$ dB. Thus, it is clear that much work remains to be done to reach the full potential of improving the sensitivity of atomic sensor via use of quantum entanglement. For the protocol proposed here, under ideal conditions, the corresponding reduction in variance would be by a factor of $54$ dB, for $N=5 \times 10^{5}$~\cite{CommentVariance}. Of course, realization of the protocol proposed here, under ideal conditions, would be difficult using the types of experimental OAT squeezing apparatus that have been implemented in various laboratories, such as those in references~\cite{Leroux1} and~\cite{Hosten}. However, it may be possible to devise alternative techniques or cavities with much higher cooperativity factors to approach the degree of improvement predicted by the protocol proposed here, as discussed in Section~\ref{Experiment}.

The rest of the paper is arranged in the following way. In Section~\ref{CRAIN_COSAIN}, we review briefly the theory of the CRAIN and the COSAIN. Section~\ref{SCAIN} provides a detailed description of the protocols employed for a SCAIN, as well as the resulting signal fringes and sensitivities. Section~\ref{Experiment} gives a brief description of the two key experimental components for implementing a SCAIN (namely, collective state detection and OAT squeezing), and a discussion about the practical challenges and limitations. In Appendix A, we discuss how the physical interpretation for the phase magnification for the SCAIN is different for different modes of operation: enhancement of the Compton frequency for rotation sensing, and enhancement of the effective two-photon wave vector for accelerometry.  In Appendix B, we present a detailed description of the Schr\"odinger Cat Atomic Clock (SCAC). Finally, in Appendix C, we provide an alphabetical list of abbreviations used in this paper.

\section{Brief review of the CRAIN and the COSAIN}
\label{CRAIN_COSAIN}
In order to illustrate clearly the mechanism for realizing the SCAIN, and the characteristics thereof, as well as to establish the notations employed in the rest of this paper, it is useful to recall briefly the relevant features of a CRAIN and a COSAIN. A CRAIN makes use of $N$ non-interacting identical three-level atoms with metastable hyperfine states $\ket{\downarrow, p_z =0}$ and $\ket{\uparrow, p_z =\hbar k}$, (where $k = k_1 + k_2$, with $k_1$ and $k_2$ being the wave numbers for the two counter-propagating beams, and $p_z$ being the $z$-component of the linear momentum), and an excited state $\ket{e}$, in the $\Lambda$-configuration, reduced to an equivalent two-level model~\cite{Shahriar}. We represent these atoms by a collective spin $\vu{J} = \sum_i^N\vu{j}_i$, where $\vu{j}_i$ represents the pseudospin-$1/2$ operator for each atom. The ensemble is initially prepared in a CSS: $\ket{-\vu{z}}\equiv\ket{E_0} = \prod_{i=1}^N\ket{\downarrow_i}$. Here, we have employed the notation that $\ket{\vu{w}}$ represents a CSS where the pseudo-spin of each atom is aligned in the direction of the unit vector $\vu{w}$. Under a pulse sequence of $\pi/2-$dark$-\pi-$dark$-\pi/2$, each atom's wavepacket first separates into two components, then gets redirected and finally recombined to produce an interference which is sensitive to any phase-difference, $\phi$ between the two paths. As an example, we consider the case of rotation where an AI gyroscope rotating at a rate $\Omega_G$ about an axis normal to the area $\Theta$ accrues a phase difference $\phi= 2\omega_c\Theta\Omega_G/c^2$ between its trajectories~\cite{Sagnac}. The effect of the overall phase shift $\phi$ due to rotation is uniformly spread throughout the interferometric sequence. However, for theoretical convenience, we introduce it in two equal parts during each of the dark zones (a justification of this approach can be found in Ref.~\cite{Shahriar2}). The final state of the atoms is given by
\begin{align}
\ket{\psi} = e^{-i\frac{\pi}{2}\hat{J}_x}e^{i\frac{\phi}{2}\hat{J}_z}e^{-i\pi\hat{J}_x}e^{-i\frac{\phi}{2}\hat{J}_z}e^{-i\frac{\pi}{2}\hat{J}_x}\ket{-\vu{z}}\nonumber\\
= \prod_{i=1}^N-\frac{1}{2}e^{-i\phi/2}((1+e^{i\phi})\ket{\downarrow_i}+i(1-e^{i\phi})\ket{\uparrow_i}).
\end{align}

In a CRAIN, $\phi$ is measured by mapping it onto the operator representing the difference in spin-up and spin-down populations: $\hat{J}_z = (\hat{N}_{\uparrow} -\hat{N}_{\downarrow})/2$, where $\hat{N}_{\uparrow} = \Sigma_i\ket{\uparrow_i}\bra{\uparrow_i}$ and $\hat{N}_{\downarrow} = \Sigma_i\ket{\downarrow_i}\bra{\downarrow_i}$. The signal, which is a measure of the population of $\ket{\downarrow}$ is, therefore, $S_{CRAIN} = J + \langle - \hat{J}_z \rangle = N\cos^{2}(\phi/2)$, where $J = N/2$. The corresponding fringe linewidth is given by $\varrho=c^2/(2\omega_c\Theta)$. The measurement process causes wavefunction collapse of the individual spins from the superposition state to $\ket{\downarrow}$, resulting in quantum projection noise in the measure of the signal~\cite{Wineland3}, $\Delta S_{CRAIN} = \Delta (-\hat{J}_z)=\sqrt{N/4}\sin(\phi)$, where $\Delta \hat{J}_z$ is the standard deviation of $\hat{J}_z$. Assuming ideal quantum efficiency, the Quantum Fluctuation in Rotation-rate (QFR) is given by $\Delta \Omega_G\bigr|_{CRAIN}=\abs*{\Delta (-\hat{J}_z)/\partial_{\Omega_G} \langle -\hat{J}_z \rangle }= c^2/2\omega_C\Theta\sqrt{N}$, where $\partial_{\Omega_G}\equiv \partial/\partial \Omega_G$.

The COSAIN differs from a CRAIN in that the measurement of the signal is done on a Dicke collective state of the ensemble, instead of a single atomic state~\cite{COSAIN}. The Dicke states are eigenstates of $\hat{J}_z$ and can be represented as $\ket{E_n, p_z = n\hbar k} = \Sigma_{k=1}^{N \choose n}P_k\ket{\downarrow^{N-n}\otimes\uparrow^n}/\sqrt{N \choose n}$, where $P_k$ is the permutation operator~\cite{CollectiveDescription}. As a result of the first $\pi/2$-pulse, the initial state $\ket{E_0, p_z = 0}$ is coupled to $\ket{E_1, p_z = \hbar k}$, which in turn is coupled to $\ket{E_2, p_z = 2\hbar k}$, and so on, all the way up to $\ket{E_N, p_z = N\hbar k}$. This causes the ensemble to split into $N+1$ trajectories. The dark zone that immediately follows imparts a phase $e^{in\phi/2}$ to $\ket{E_n}$. At this point, the $\pi$-pulse generates a flip in the individual spins, causing $\ket{E_n}$ to become $\ket{E_{N-n}}$, and vice versa. The second dark-zone lends a phase $e^{i(0.5N-n)\phi}$ to $\ket{E_n}$. The mathematical derivation of this mechanism is discussed in detail in Ref.~\cite{COSAIN}. The last $\pi/2$-pulse causes each of the collective states to interfere with the rest of the states. The COSAIN can, thus, be viewed as an aggregation of interference patterns due to ${N+1}\choose 2$ interferometers working simultaneously. 

The narrowest constituent signal fringes are derived from interferences between states with the largest difference in phase, i.e. $\ket{E_0}$ and $\ket{E_N}$. The width of this fringe is $\varrho/N$. The widths of the rest of the signal components range from $\varrho$ to $\varrho/(N-1)$. The signal, which is the measure of population of $\ket{E_0}$, is the result of the weighted sum of all the pairwise interferences with this state. This is detected by projecting the final state of the ensemble, $\ket{\psi}$ on $\ket{E_0}$. Thus, $S_{COSAIN} = \langle\hat{G}\rangle=\cos^{2N}(\phi/2)$, where $\hat{G} \equiv\ket{E_0}\bra{E_0}$. The quantum projection noise is the standard deviation of $\hat{G}$, given by $\Delta S_{COSAIN} = \cos^{N}(\phi/2)\sqrt{1-\cos^{2N}(\phi/2)}$. The QFR of the COSAIN is thus, $\Delta \Omega_G\bigr|_{COSAIN} = \abs*{\Delta \hat{G}/\partial_{\Omega_G} \langle \hat{G}\rangle}$. Under quantum noise limited operation, this equals $(\Delta \Omega_G\bigr|_{CRAIN}/\sqrt{N})|\sqrt{\sec^{4J}(\phi/2)-1}/\tan(\phi/2)|$. Therefore, for $\Omega_G \rightarrow 0$, the rotation sensitivity of the COSAIN is same as that of a CRAIN, which is the SQL, assuming all the other factors remain the same. One way of surpassing the SQL is to suppress the contribution of the constituent fringes broader than $\varrho/N$. This is precisely what happens in the SCAIN, which makes use of squeezed spin state (SSS) of the ensemble: $\ket{\psi_e}=e^{-i\mu J_z^2}\ket{\vu{y}}$, where $\mu$ is the squeezing parameter, and $\vu{y}$ is the quantum state produced by the first $\pi/2$-pulse.

\section{Schr\"odinger Cat Atomic Interferometer}
\label{SCAIN}
The SCAIN can be operated under two different protocols, which differ by the choice of the axis around which we apply the rotation that maximizes the degree of observed squeezing. In one case (Protocol A), the rotation is around the $\vu{x}$ axis while in the other (Protocol B), the rotation is around the $\vu{y}$ axis.

\subsection{Protocol A}

\begin{figure}[h]
\includegraphics[scale=0.32]{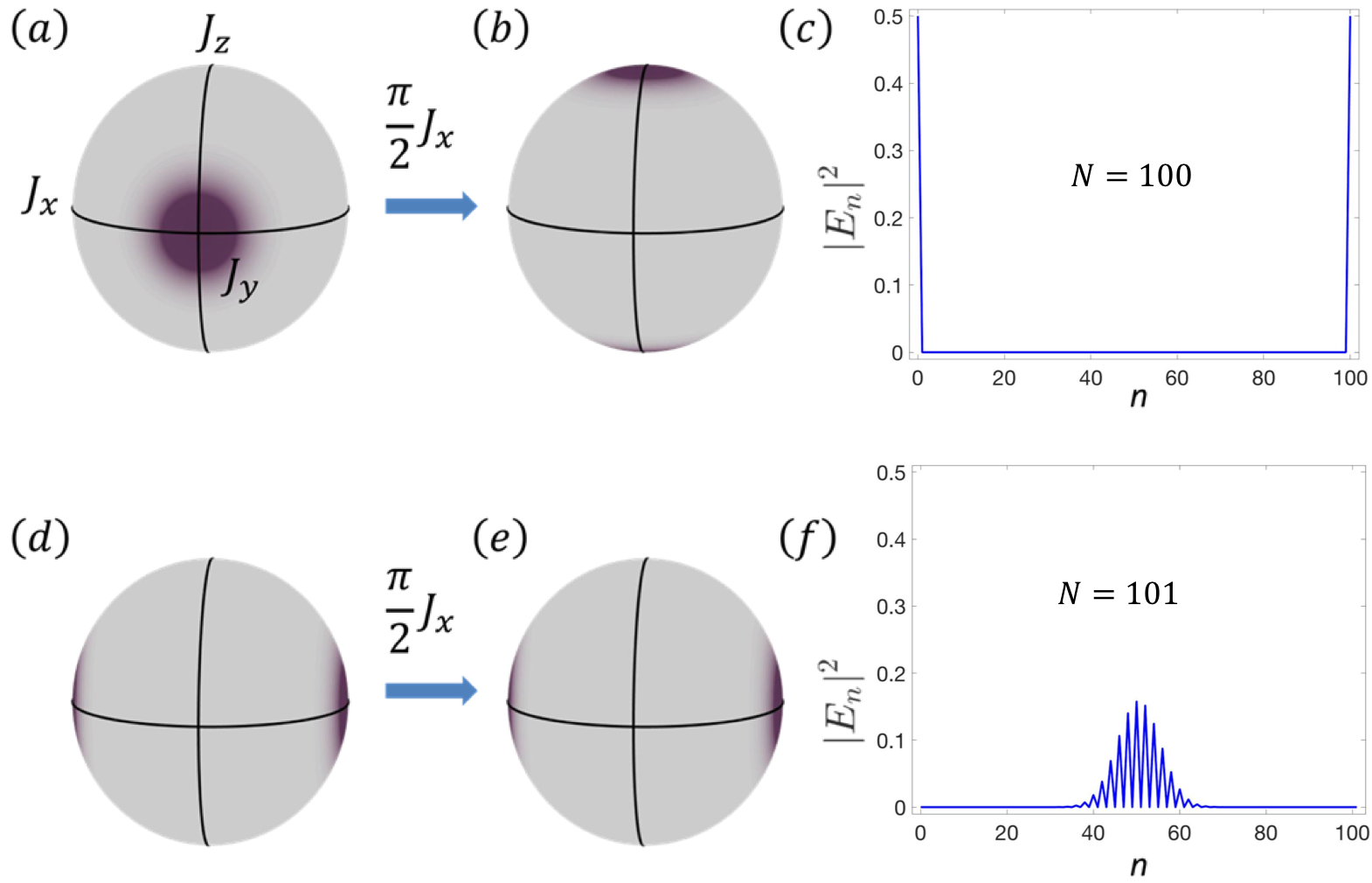}
\caption{Illustration of the SCAIN scheme for Protocol A. For even $N$: (a) For $\mu=\pi/2$, the Husimi quasi probability distribution (QPD) is split into two circular components located on the opposite faces normal to the $\vu{y}$ axis of the Bloch sphere. (a$\rightarrow$b) The QPD of the SSS state ($\ket{\psi_e}$) is rotated by $\pi/2$ about the $\vu{x}$ axis to yield the Schr\"odinger cat state; note the components on both top and bottom of the Bloch sphere in (b). (c) Distribution of collective states in the rotated SSS, showing 50$\%$ in state $\ket{E_0}$ and 50$\%$ in state $\ket{E_N}$. For odd $N$: (d) For $\mu= \pi/2$, the QPD is split into two circular components located on the opposite faces normal to the $\vu{x}$ axis of the Bloch sphere. (d$\rightarrow$e) rotation about $\vu{x}$ axis does not transform the SSS. (f) Distribution of collective states in the rotated SSS. These results also hold for the case of atomic clocks, as described in Appendix B.}
\label{fig:1}
\end{figure}

We first consider Protocol A, focusing initially on the special case where the squeezing parameter $\mu$ is $\pi/2$, as illustrated in Fig.~\ref{fig:1}, with the case of an arbitrary value of $\mu$ to be discussed later. The OAT spin squeezing effect is achieved by applying the squeezing Hamiltonian, $H_{OAT} =\hbar \chi J_z^2$, for a duration of time $\tau$ such that $\mu = \chi\tau$. For even $N$, $H_{OAT}$ transforms $\ket{\vu{y}}$ to $\ket{\psi_e}=(\ket{\vu{y}} -\eta \ket{-\vu{y}})/\sqrt{2}$, where $\eta \ = i(-1)^{N/2}$, representing a phase factor with unity amplitude. Rotating $\ket{\psi_e}$ by an angle of $\nu = \pi/2$ about the $\vu{x}$ axis yields the Schr\"odinger cat state $\ket{\psi_{SC}} = (\ket{E_0} + \eta \ket{E_N})/\sqrt{2}$. At the end of the intermediate dark$-\pi-$dark sequence, the state of the ensemble is $e^{i\phi J_z/2}e^{-i\pi J_x}e^{-i\phi J_z/2}\ket{\psi_{SC}} = (e^{iN\phi/2}\eta \ket{E_N} + e^{-iN\phi/2}\ket{E_0})/\sqrt{2}$. As discussed above, the interference between states with a phase difference $N\phi$ produces signal fringes narrowed by a factor of $N$. To measure $\phi$, we seek to undo the effect of squeezing on the system. This is accomplished in two steps. First, we apply another rotation $\nu = \pi/2$ (rather than $-\pi/2$, as noted earlier, due to the state inversion caused by the $\pi$-pulse) about the $\vu{x}$ axis. Thereafter, the untwisting Hamiltonian, $-H_{OAT}$ is applied. Finally, the last $\pi/2$ pulse is applied to catalyze interference between the resulting states. The signal arising from this interference depends on $\phi$ as $S_{SC} =\langle\hat{G}\rangle= \sin^2(N\phi/2)$.

When $N$ is odd, initial squeezing produces $\ket{\psi_e}=(\ket{\vu{x}} +\zeta \ket{-\vu{x}})/\sqrt{2}$, where $\zeta=i(-1)^{(N+1)/2}$, also representing a phase factor with unity amplitude. For $\phi = 0$, the sequence $e^{-i\nu J_x}e^{i\phi J_z/2}e^{-i\pi J_x}e^{-i\phi J_z/2}e^{-i\nu J_x}$ only causes an identical phase change in each of these states. Application of the unsqueezing Hamiltonian, $-H_{OAT}$ then restores the system to $\ket{\vu{y}}$, and the final $\pi/2$ pulse places the system in the $\ket{\vu{z}}$ state, which is the same as the collective state $\ket{E_N}$. Since we detect the collective state $\ket{E_0}$, the whole sequence thus generates a null signal. For reasons that are not manifestly obvious due to the complexity of the states, but can be verified via simulation, the same conclusion holds for an arbitrary value of $\phi$. Over repeated measurements, the probability of $N$ being even or odd is equal. Thus, for $M$ trials, the average signal of the SCAIN in this regime is $S_{SC} = M\sin^2(N\phi/2)/2$. The associated quantum projection noise is $\Delta S_{SC} = \sqrt{M/2}\sin(N\phi)$. The QFR is thus, $\Delta\Omega_G = c^2/\sqrt{2M}N\omega_C\Theta$, which is a factor of $\sqrt{2}$ below the HL.

\subsection{Protocol B}
Next we consider Protocol B. In this protocol, the rotation is always around $\vu{y}$ axis while the rotation angle $\nu$ is chosen so as to maximize (right after the squeezing interaction) the fluctuations along $\vu{z}$ axis. For a given value of $N$, $\nu$ increases with $\mu$, reaching a maximum value of $\pi/2$ at $\mu=\mu_0$ (a typical value of $\mu_0$ is $0.095\pi$ for $N=200$, for example). Once the SSS is optimally aligned, the usual dark$-\pi-$dark sequence follows. To undo the effect of the squeezing, we first apply another rotation $\nu$ about $\vu{y}$ axis, and then apply $-H_{OAT}$. Finally, the last $\pi/2$ pulse is applied to catalyze interference between the two paths of the interferometer. 

\begin{figure}[h]
\includegraphics[scale=0.40]{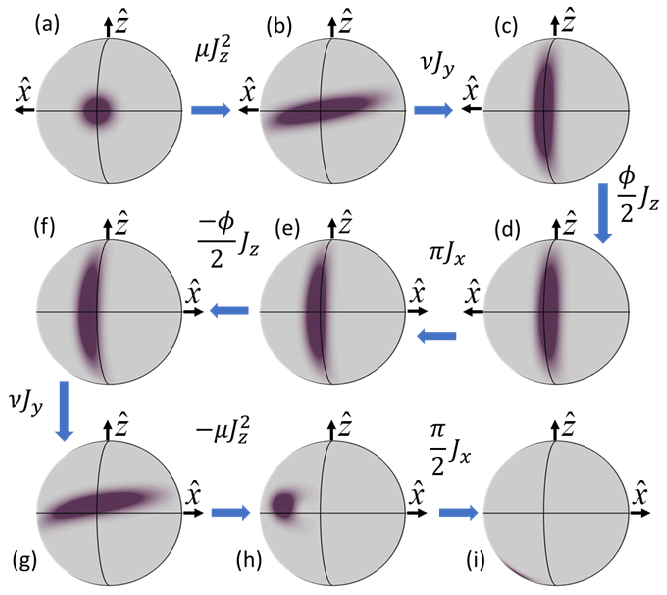}
\caption{The evolutions of QPD for Protocol B with $\mu<\pi/2$. The initial CSS $\ket{\vu{y}}$ (a) evolves under $H_{OAT}$ to (b) which is then rotated by an angle $\nu$ (b$\rightarrow$c) so as to maximize the fluctuations along $\vu{z}$. (d) The first dark zone imparts a phase $\phi/2$. (e) The Bloch sphere is rotated to show the other face where the SSS is situated after the $\pi$ pulse. (f) The second dark zone imparts an additional $\phi/2$ phase. (d$\rightarrow$g) The spins are unrotated by the same angle $\nu$ and then (h) unsqueezed, by applying the inverse of $H_{OAT}$. (i) The final $\pi/2$ pulse causes interference between the two paths of the interferometer.}
\label{fig:2}
\end{figure}

In Fig.~\ref{fig:2}, we show the QPD evolutions for Protocol B with $\mu< \pi/2$. After the first $\pi/2$-pulse, the system is in the CSS $\ket{\vu{y}}$, as shown in Fig.~\ref{fig:2}~(a). Following the application of the squeezing interaction, the quantum fluctuations are twisted in the $x$-$z$ plane, as depicted in Fig.~\ref{fig:2}~(b). We then apply a rotation around the $\vu{y}$ axis by an angle $\nu$ which is chosen so as to maximize the fluctuations along the $\vu{z}$ axis, as illustrated in Fig.~\ref{fig:2}~(c). For a given value of $N$, $\nu$ increases with $\mu$, reaching a maximum value of $\pi/2$ at $\mu=\mu_0$ (for $N=200$, $\mu_0=0.095\pi$). Once the SSS is optimally aligned, the usual dark$-\pi-$dark sequence follows, where the first and second dark zones each impart a phase of $\phi/2$ to the SSS, while the $\pi$-pulse inverts the states. These are shown in Fig.~\ref{fig:2}~(d)-(f). To undo the effect of the squeezing, we first apply another rotation $\nu$ about $\vu{y}$ axis, and then apply $-H_{OAT}$, as depicted in Fig.~\ref{fig:2}~(g)-(h). Finally, the last $\pi/2$ pulse is applied to catalyze interference between the two paths of the interferometer, as shown in Fig.~\ref{fig:2}~(i).

\begin{figure}[h]
\includegraphics[scale=0.36]{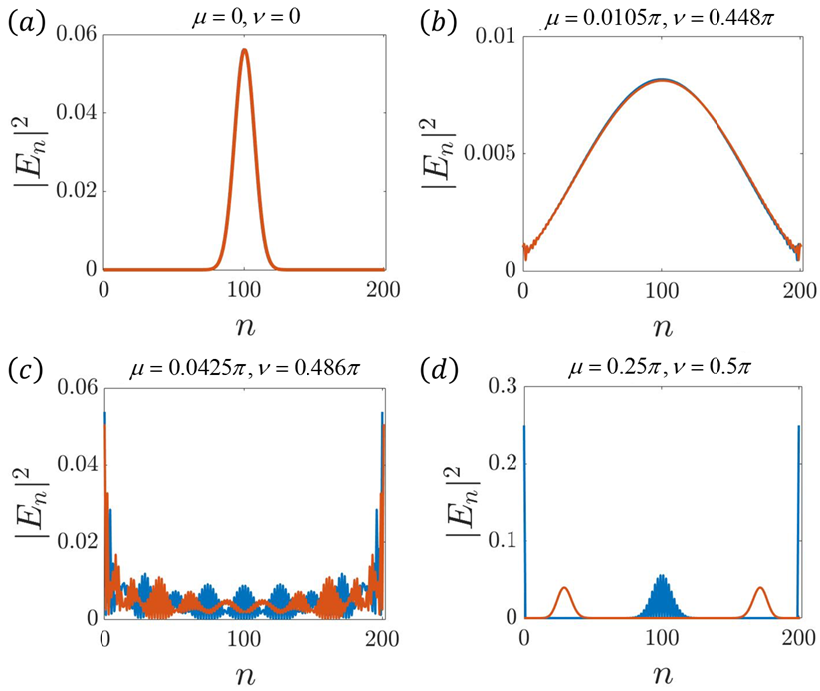}
\caption{Collective state population distributions right after the squeezing interaction for different values of $\mu$, under Protocol B. Both even (blue line) and odd (red line) values of $N$ are considered. These results also hold for the case of atomic clocks, as described in Appendix B.}
\label{fig:3}
\end{figure}

In Fig.~\ref{fig:3}, we show the collective state population distributions right after the squeezing interaction for different values of $\mu$, under Protocol B. For $\mu=0$, the SSS has the same binomial distribution of the collective states as in the original CSS, as depicted in Fig~\ref{fig:3}~(a). As $\mu$ increases, the distribution begins to flatten out, as shown in Fig~\ref{fig:3}~(b). When $\mu$ becomes large enough, the distribution starts to invert, and the relative proportion of the extremal states increases. However, the exact state distribution depends on the parity of $N$, as demonstrated in Fig~\ref{fig:3}~(c). At $\mu=\pi/4$, the distribution is trimodal for even values of $N$, as depicted by the blue line in Fig.~\ref{fig:3}~(d). On the other hand, for odd values of $N$, the distribution is bimodal, as shown by the red line in Fig.~\ref{fig:3}~(d).

%

\subsection{Signal fringes under the two protocols}
\begin{figure}[h]
\includegraphics[scale=0.3]{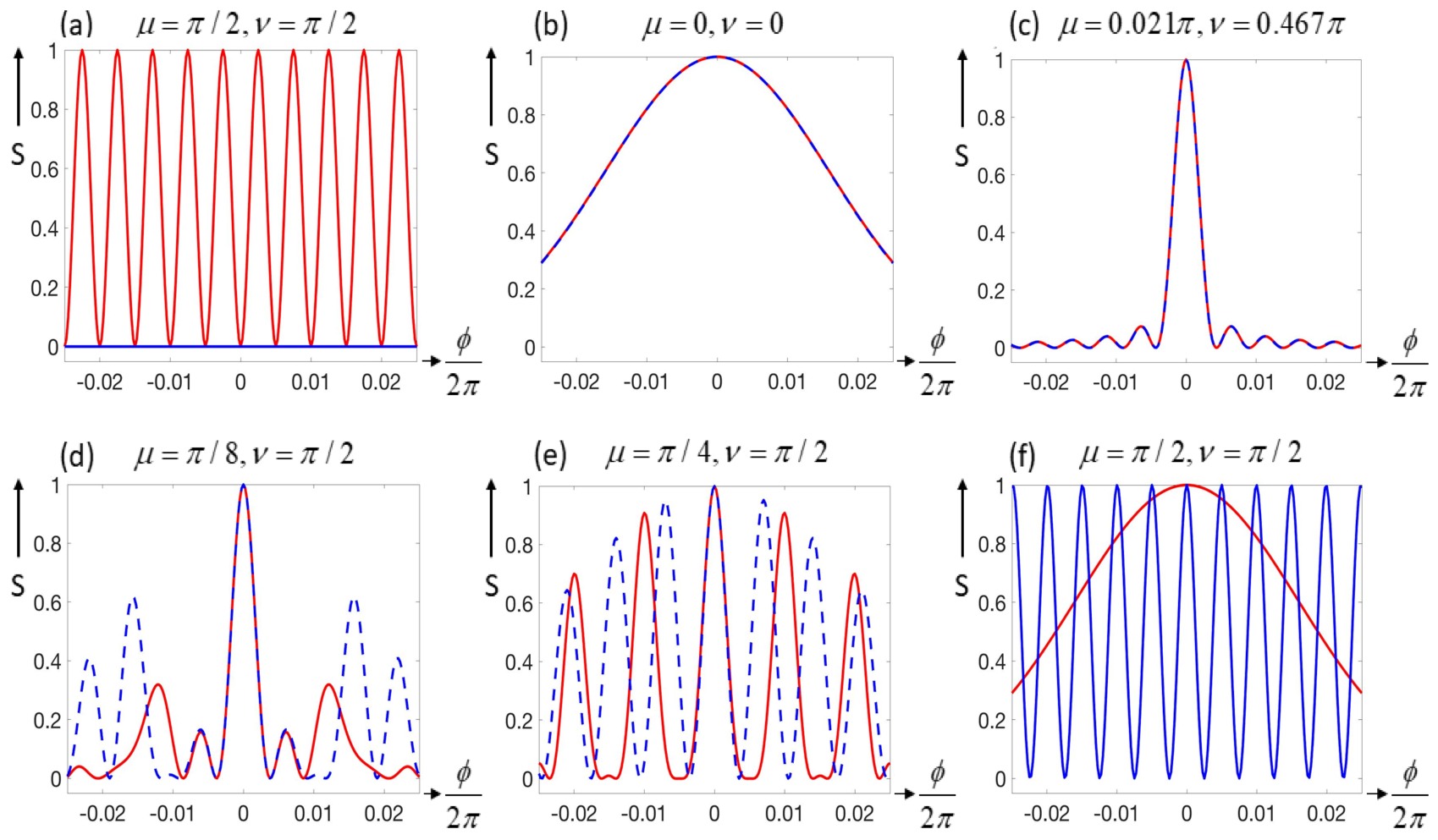}
\caption{Signal fringes for various values of $\mu$. $N = 200$ is indicated by red lines, $N = 201$ by blue (dashed or solid) lines. Figure (a) employs Protocol A, while figures (b)-(f) employ Protocol B. The phase span is 1/20-th of $2\pi$; as such, we see $10$ red fringes in figure (a), thus demonstrating a factor of $N$ reduction in the width of fringes for Protocol A.}
\label{fig:4}
\end{figure}

In Fig.~\ref{fig:4}~(a), we show the signal fringes produced using Protocol A, for the special case of $\mu=\pi/2$. As described earlier, in this case, we get a purely sinusoidal fringe pattern for even values of $N$, and a null signal for odd values of $N$. The averaged signal, therefore, is also purely sinusoidal. The width of these fringes is a factor of $N$ narrower than what is observed in a CRAIN. It should be noted that the phase factors $\eta$ and $\zeta$ (as defined for the superpositions of collective states generated via the first application of OAT squeezing under Protocol A described above) depend, respectively, on the super even parity (SEP), representing whether $N/2$ is even or odd, and the super odd parity (SOP), representing whether $(N+1)/2$ is even or odd. However, in each case, the shapes of the fringes as well as the values of QFR, are not expected to depend on the value of SEP and SOP, as we have verified explicitly.

The signal fringes under Protocol B are illustrated in Fig.~\ref{fig:4}~(b)-(f), for various values of $\mu$. The red lines are for even values of $N$, and the blue (dashed or solid) lines for odd values of $N$. For different values of $\mu$ (except for $\mu=\pi/2$), the central fringe as a function of $\phi$ is essentially identical for both odd and even values of $N$. Thus, for $M$ trials, the average signal is independent of the parity of $N$ for the central fringe, which is the only one relevant for metrological applications. In contrast, the non-central fringes, averaged over the odd and even cases, have different shapes, heights and widths. However, the central fringe always has full visibility, and its width first decreases sharply with increasing values of $\mu$, and then saturates at $\mu=\mu_0$. Consequently, the fluctuations in rotation sensitivity plummets, attaining the minimum value $\Delta \Omega_G|_{SCAIN} = e^{1/3}c^2/2\sqrt{M}N\omega_C\Theta$, at $\mu=\mu_0$.

For the limiting case of $\mu = \pi/2$, Protocol B produces very different results for odd and even values of $N$. Specifically, for odd values of $N$, this protocol produces uniform fringes, each with a width that is a factor of $N$ narrower than what is observed in a CRAIN, thus yielding HL sensitivity. In this case, the ideal Schr\"odinger Cat state is realized, in a manner analogous to what we described above for Protocol A (with $\mu=\pi/2$). For odd values of $N$, this protocol also produces uniform fringes, but each with a width that is the same as that observed for COSAIN (which is a factor of $\sqrt{N}$ narrower than what is observed in a CRAIN), thus yielding SQL sensitivity. The average of these two signals, for many repeated measurements, would produce a sensitivity that, for large $N$, is lower than the HL by a factor of $\sqrt{2}$~\cite{Shahriar2}. In addition, due to the mixing of the suboptimal signal contributed by the instances corresponding to even values of $N$, Protocol B, even for $\mu=\pi/2$, is not well-suited for experiments aimed at studying the effects of gravity on clear superposition of just two macroscopic states~\cite{Diosi, Penrose1, Penrose2, Penrose3},  and realizing a matter-wave clock with very high Compton frequency~\cite{Lan}.

\subsection{QFR$^{-1}$ under the two protocols}

\begin{figure}[h]
\includegraphics[scale=0.4]{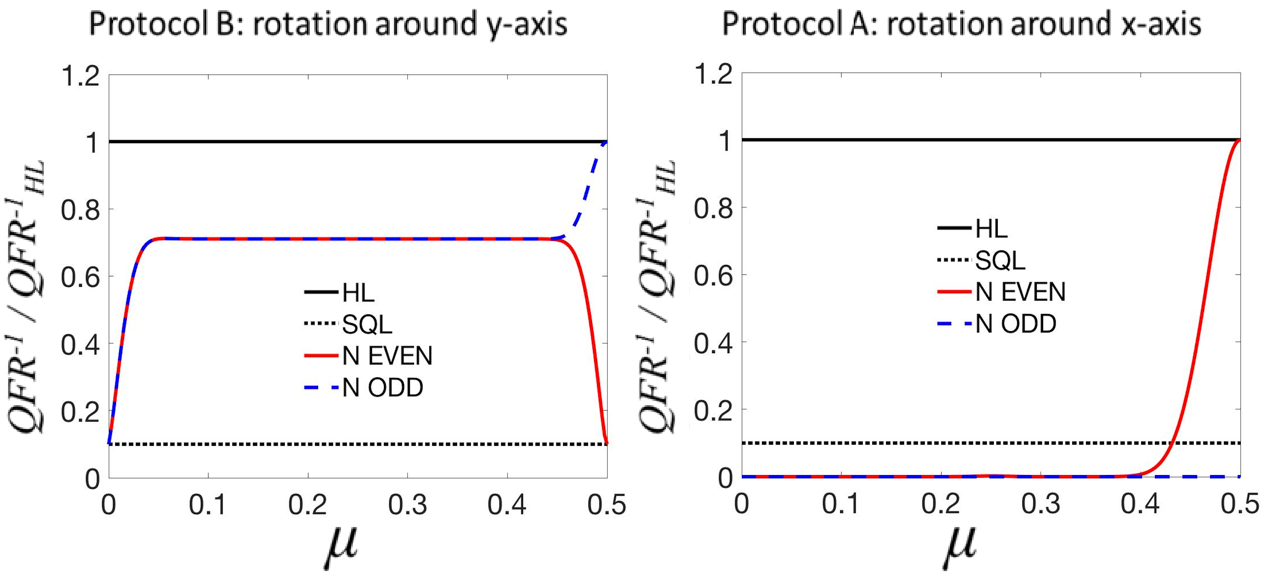}
\caption{QFR$^{-1}$ of SCAIN as a function of the squeezing parameter, $\mu$, normalized to the HL for $N = 100$. Note that, for this value of $N$, the HL corresponds to a gain in sensitivity by a factor of 10 compared to the SQL. Horizontal lines indicate the HL (black solid) and the SQL (black dashed). The dashed blue lines correspond to odd value of N ($N = 101$) and the red lines correspond to even value of N ($N = 100$). The left(right) panel shows the results for Protocol B(A).}
\label{fig:5}
\end{figure}

In Fig.~\ref{fig:5}, we summarize the results for both protocols, for squeezing parameters ranging from $\mu=0$ to $\mu=\pi/2$. Here, we show the inverse of the QFR, normalized to the same for the HL for $N = 100$, as a function of $\mu$. Horizontal lines indicate the HL (black solid), and the SQL (black dashed), where for $N=100$, the HL corresponds to a gain in sensitivity by a factor of $10$ compared to the SQL. The dotted blue lines correspond to odd value of N ($N=101$) and the red lines correspond to even value of N ($N = 100$). The left panel shows the result of using Protocol B. The value of QFR$^{-1}$ increases monotonically, reaching a peak value at $\mu=\mu_0$, and then remains flat until getting close to $\mu=\pi/2$, with virtually no difference between the odd and even values of $N$, as discussed in detail earlier. Near $\mu= \pi/2$, the value of QFR$^{-1}$ begins to diverge, reaching the HL(SQL) for odd(even) values of $N$ at $\mu=\pi/2$. The right panel shows the result of using Protocol A. At $\mu=\pi/2$, QFR$^{-1}$ is at the HL for even values of $N$, and vanishes for odd values of $N$. For $\mu < \pi/2$, the amplitude of the signal for even values of $N$ decreases rapidly, with corresponding decrease in the value of QFR$^{-1}$. It should be noted that a vanishing value of QFR$^{-1}$ is due simply to the vanishing of the signal itself.

\section{Experiment Considerations for Realizing the SCAIN}
\label {Experiment}

In this section, we describe the experimental steps envisioned for realizing the SCAIN, and discuss potential limitations. The basic protocol is akin to that employed for the CRAIN, with the addition of auxiliary rotations, one axis twist (OAT) squeezing and collective state detection (CSD). In what follows, we first summarize briefly the experimental approach for OAT squeezing and CSD that are well-suited for the SCAIN. This is followed by a discussion of the complete protocol for the SCAIN. We discuss both Protocols A and B, but limit the description to the case of $\mu=\pi/2$. The case for $\mu<\pi/2$ can be easily inferred from this discussion. 

There are several experimental schemes for realizing one-axis-twist squeezing~\cite{Takeuchi,Schleier,Leroux1,Leroux2,Hosten,Antisqz,Antisqzexpt,Wang,Liu,Gil}. For concreteness, we consider here the approach based on cavity feedback dynamics~\cite{Schleier,Leroux1,Leroux2,Hosten,Antisqz,Antisqzexpt}. In this approach, a probe is passed through a cavity, at a frequency that is tuned halfway between the two legs of a $\Lambda$ transition in which the spin-up and spin-down states are coupled to an intermediate state. The cavity is tuned to be below resonance for the probe. The energy levels of the spin-up and spin-down states are light shifted due to the probe, in opposite directions. The resulting dispersion shifts the cavity resonance frequency by an amount that is proportional to $J_z$, the z-component of the total spin for all atoms. The intra-cavity probe intensity changes linearly with this cavity shift, since it is on the side of the resonance, thus affecting the light-shifts. The net result is an energy shift for all the atoms that is proportional to the square of $J_z$, so that the interaction Hamiltonian can be expressed as $H_{OAT} =\hbar \chi J_z^2$, where $\chi$ is a parameter that determines the strength of the squeezing process. Changing the sign of the cavity detuning reverses the sign of the Hamiltonian, thus producing unsqueezing. 

The collective state detection technique is detailed in section IV of Ref.~\cite{COSAIN}, where a null-detection scheme is employed to measure population of one of the extremal Dicke collective states. The probe is one of the two counter-propagating Raman beams, which induces Raman transitions within the atomic ensemble unless it is in the desired extremal collective state. As a result, there will be photons emitted corresponding to the other leg of the Raman transition. The probe and the emitted photons are combined and sent to a high speed detector, which produces a dc signal along with a beat signal. This beat signal is at the same frequency as that of the signal produced by the frequency synthesizer (FS) that drives the Acousto-Optic Modulator (AOM), for example, used to generate the beam that excites one leg of the Raman excitation from the beam that excites the other leg of the Raman excitation, but with a potential difference in phase. To extract the amplitude, the beat signal is bifurcated and one part is multiplied by the FS signal, while the other is multiplied by the FS signal phase shifted by $90$ degrees. The signals are then squared before being recombined and sent through a low-pass filter (LPF) to derive a dc voltage. This dc voltage is proportional to the number of scattered photons. A lower limit (ideally zero) is set for the voltage reading, and any value recorded above it indicate the presence of emitted photons. If no photon is emitted, the voltage will be at or below the limit, indicating that the ensemble is in the desired extremal collective state; otherwise at least one photon will be emitted and the ensemble will be in a combination of other collective states. This process is then repeated many times for a given value of $\phi$. The fraction of events where no photons are detected will correspond to the signal for this value of $\phi$. This process is then repeated for several values of $\phi$, producing the signal fringe.

For the complete SCAIN experiment, we assume that the source atoms, are caught in a magneto-optic trap (MOT), followed by polarization gradient cooling and evaporative cooling, to a temperature of about $\SI{0.5}{\micro\kelvin}$, with a phase-space density less than what is required for Bose-Einstein Condensation (BEC). The atoms are then pushed out, forming a sequential beam of N atoms in each sequence. An initial (counter-propagating) Raman pulse, corresponding to a rotation of $\pi/2$ around the x-axis, splits each atom, originally in the spin-down state, into an equal super-position of spin-up and spin-down states. The atoms then pass through a transverse ring cavity set up for OAT squeezing. The squeezing process is carried out for a duration corresponding to $\mu=\pi/2$, followed by an auxiliary rotation (produced by another pair of Raman beams) by an angle of $\pi/2$ around the x-axis. This creates the SC state, as a superposition of two extremal Dicke collective states: one in which all atoms are in the spin-down state, and another in which all atoms are in the spin-up state. The two components in the SC state get spatially separated during the first dark zone evolution. This is followed by another Raman pulse which produces a rotation of $\pi$ around the x-axis. This pulse redirects the velocities of the two components. After the second dark zone, another Raman pulse is applied for a duration that produces a rotation of $\pi/2$ around the x-axis. This is followed by an unsqueezing pulse, of duration corresponding to $\mu=-\pi/2$, which is produced by sending the atoms through a second transverse ring cavity, with a cavity detuning that is equal and opposite to the one applied in the first cavity. After the unsqueezing, the final $\pi/2$ rotation around the x-axis, produced by another Raman pulse, causes the two paths to interfere. The collective state detection process is then used to determine the population of the atoms in the collective state in which all the atoms in the spin-down state, representing the signal for the SCAIN, under Protocol A and the limiting case of $\mu=\pi/2$.  

For implementing Protocol B, for $\mu=\pi/2$, the basic sequence is the same as what is described above, with the following modifications. Note that, in the sequence described above, there are five different pairs of Raman beams; three of these are used for the conventional pulse sequences necessary for a CRAIN, while the other two are used for auxiliary rotations.  In the case of Protocol B, the auxiliary rotations are around the y-axis. The phase of the beat signal between the two frequencies employed for Raman excitation determines the axis of rotation. Thus, this phase for the two pairs of Raman beams used for the auxiliary rotations has to be shifted by $90$ degrees compared to the same for the three pairs of Raman beams used for the CRAIN pulse sequence. To see how this phase shift can be produced, we note that (as also mentioned in the discussion for the CSD above) for each pair of Raman beams, we start with a laser beam at a frequency that excites one leg of the $\Lambda$ transition. The second laser frequency, which excites the other leg of the $\Lambda$ transition, is produced by shifting the frequency of a piece of the first laser beam by passing it through an AOM, for example. The frequency that drives the AOM is generated from an FS. Thus, to generate the phase shift needed for Protocol B, we lock the difference between the phase of the FS used for the auxiliary Raman beams and that of the FS used for the CRAIN Raman beams to a value of $90$ degrees. As a result, the auxilary Raman beams will produce rotations of $\pi/2$ around the y-axis, as needed for Protocol B. 

To elucidate potential practical limitations in implementing the SCAIN protocol experimentally, as envisioned above, consider first the situation where the OAT squeezing and unsqueezing processes are ideal.  In that case, the relevant issues pertain to the potential imperfections in generating the ideal collective states. In references~\cite{COSAIN} and~\cite{CollectiveDescription}, we discussed the issues that are relevant in this context, and how these issues may limit the performance of the COSAIN. Essentially the same issues are expected to constrain the performance of the SCAIN. In what follows, we summarize the findings of the analysis presented in these two references~\cite{COSAIN, CollectiveDescription}, in the context of the SCAIN, using $^{87}$Rb atoms for specificity. First, we noted that for a Raman excitation based atomic interferometer (such as the COSAIN and the SCAIN), the collective states must be defined in a manner so that the spin-down state represents the atom being in the ground state of the internal energy, and in a momentum eigenstate of the center-of-mass (COM) motion, and the spin-up state represents the atom being in a higher-energy but metastable internal state, and in another momentum eigenstate of the COM motion. Since the atom is in a wavepacket with respect to the COM motion, the spin-down state, for example, is in a superposition of momentum eigenstates. Similarly, the spin-up state is also in a superposition of momentum eigenstates, even if we assume that the two-photon recoil imparted by the Raman beams is exactly the same for each atom. In Section $4$ of reference~\cite{CollectiveDescription}, we addressed this issue explicitly, and showed that if the effective Rabi frequency of the off-resonant Raman transition (i.e., the Raman Rabi frequency) is much larger than the Doppler shift due to the COM momentum of each of the constituent plane waves in the ground state wavepacket, then the description of the semi-classical collective states (which ignores the COM motion), as employed here and in virtually all descriptions of collective states in the literature, remains valid.  For the temperature of $\SI{0.5}{\micro\kelvin}$ mentioned above for the SCAIN, it should easily be possible to realize an effective Rabi frequency large enough to satisfy this condition.  

Second, we considered the effect of the variations in the intensity profiles of the laser beams, which in turn cause variations in the Raman Rabi frequency. The effect of this inhomogeneity can be mitigated by increasing the ratio, $\rho$, of the diameter of the Raman beams to the diameter of the atomic cloud. For $\rho = 10$, the upper bound of the useful value of $N$ was found to be $\sim$ $1.2 \times 10^5$.  Third, we considered the effect of the velocity distribution, which causes variations in the two-photon detuning. We found that at a temperature of $\SI{0.5}{\micro\kelvin}$, this inhomogeneity limits the useful value of $N$ to $\sim$ $2 \times 10^4$. The useful value of $N$ can, in principle, be increased further by using colder atoms, as long as the phase space density is kept below the value at which BEC occurs.

Fourth, we considered the effect of spontaneous emission, since there is a small fraction of atoms in the intermediate state during the application of the Raman pulses. A proper analysis of the effect of spontaneous emission would require the use of a density matrix based model in the basis of the collective states. Coherent excitation of the atoms only populates the $(N+1)$ symmetric collective states~\cite{Dicke, Arecchi, CollectiveDescription}. However, the total number of collective states, which include the asymmetric ones, is $2^{N}$, the size of the Hilbert space for $N$ two-level atoms~\cite{CollectiveDescription}. All of these states must be taken into account when considering the effect of spontaenous emission, which can couple to both symmetric and asymmetric states. Thus, even for a modest number of $N$ that would be relevant for a SCAIN, such an analysis is intractable (as also noted in the supplement of reference ~\cite{Antisqz}). For large $N$, one must rely on experiments to determine the degree to which the generation and detection of the SC state would be affected by the spontaneous emission process during Raman excitations. However, it should be noted that the effect of spontaneous emission can be suppressed to a large degree by simply increasing the optical detuning while also increasing the laser power. This is the approach used, for example, in reducing the effect of radiation loss of atoms in a far-off resonant trap (FORT). 

Finally, we considered the effect of the fluctuations in the value of $N$. In our discussion for the SCAIN above, we have already assumed an averaging over odd and even parities of atoms, for the case where atoms are released from a trap. In addition, one must consider the fact that the mean value of $N$ itself is expected to fluctuate in this case. As we have shown in reference~\cite{COSAIN}, such a fluctuation would simply cause of the width of the fringes due to interference between the extremal collective state to deviate from the ideal value, which is a factor of $N$ narrower than the fringes in a CRAIN. Thus, for example, a fluctuation in the value of $N$ by $1\%$ would cause an $\sim$ $1\%$ fluctuation in the value of the QFR$^{-1}$. 

We also note that, in general, these constraints are much less stringent for the case of the Schr\"odinger Cat Atomic Clock (SCAC), as described in Appendix II, which is based on the use of co-propagating Raman beams or a direct microwave excitation. For example, as shown in reference~\cite{COSAC}, for the case of co-propgating Raman excitation, the velocity distribution limits the useful value of $N$ to $\sim$ $2 \times 10^6$ even for a temperature as large as $\SI{138}{\micro\kelvin}$. Similarly, for $\rho = 20$, the effect of laser intensity inhomegeneity limits the useful value of $N$ to $\sim$ $2 \times 10^6$ as well. 

Consider next the challenge in implementing the idealized OAT process as envisioned above. In the experiments done to date, employing OAT squeezing, such as those in references~\cite{Leroux1} and~\cite{Antisqzexpt}, the typical maximum value of the squeezing parameter, $\mu$, is $\sim 0.01$. To the best of our knowledge, the highest value of $\mu$, $\sim 0.0125$, was observed in references~\cite{Leroux1}. For the protocol proposed here, the ideal value of $\mu$ that produces the Schroedinger Cat states is $\pi /2$. Under ideal conditions, this value can be achieved by increasing the duration of the squeezing pulse, or increasing its intensity, for example. However, because of the various non-idealities, as discussed in detail in several papers, including the supplement of reference~\cite{Antisqz}, it is clear that, for the current experimental implementations, the quantum state after such a strong degree of squeezing interaction would be severely degraded. The non-idealities that degrade the quantum state of the ensemble include the effect of back-action due to the cavity decay, as well as due to spontaneous emission that causes spin-flips. As noted in reference~\cite{Antisqz}, the effect of both of these non-idealities can be suppressed by increasing the cooperativity parameters for the cavity (e.g., by making the cavity mode small enough so that the vacuum Rabi frequency would be much stronger than both the cavity decay rate and the rate of spontaneous emission).

However, it should be noted that, for the OAT Squeezing based protocols that have been considered so far, the maximum useful squeezing is produced for very small values of $\mu$, of the order of $\sim 0.01$ for $\sim$ half a million atoms. Because of other non-idealities, such as poor quantum efficiency of detection, the currently achieved values of squeezing are not limited by the values of $\mu$. Furthermore, under conventional protocols employing OAT Squeezing, the Hussimi quasi probability distribution begins to get distorted when $\mu$ is increased beyond $\sim 0.01$, and the magnitude of the normalized Bloch vector starts getting smaller than unity. In fact, the factor of improvement in sensitivity due to squeezing drops to unity and even less than unity for $\mu$ far below the value of $\pi/2$. As such, experimental efforts to date have been focused on eliminating these non-idealities, instead of constructing apparatuses that would increase the cooperativity parameter significantly, or exploring new schemes for OAT squeezing that would be more robust again dephasing processes. 

An important point of this paper is to show that there is a regime of OAT squeezing (namely when $\mu = \pi/2$) that produces ideal quantum states, such as a superposition of two extremal Dicke collective states, without distortion and any reduction in the amplitude of the Bloch vector. Previously, such a state has only been demonstrated for very few ions (such as in reference~\cite{Blatt}). For a very large value of $N$, the number of particles, generating such a state requires knowing the parity of $N$. Therefore, no previous study has been carried out to show how to construct a protocol under which the Heisenberg Limit (within a factor of $\sqrt 2$) can be reached even when averaging over both parities of $N$. This is the main point of this paper. We believe that the results shown in this paper would identify the need for, and generate an interest in, developing improvements in experimental implementation of OAT squeezing in a manner that makes it possible to reach a value of $\mu = \pi/2$, without significant degradation of coherence.

\section{Conclusion}
In this article, we have presented a protocol for an atom interferometer that reaches the Heisenberg Limit (HL), within a factor of $\sim$ $\sqrt{2}$, via collective state detection and critical tuning of one axis twist spin squeezing. It generates a Schr\"odinger cat state, as a superposition of two collective states.  When this Schr\"odinger Cat Atom Interferometer (SCAIN) is configured as a gyroscope, the interference occurs at an ultrahigh Compton frequency, corresponding to a mesoscopic single object with a mass of $Nm$, where $N$ is the number of particles in the ensemble, and $m$ is the mass of each particle.  The signal for the SCAIN is found to depend critically on the parity of $N$. We present two variants of the protocol. Under Protocol A, where the auxiliary rotation occurs around the x-axis, the fringes are narrowed by a factor of $N$ for one parity, while for the other parity the signal is zero. Under Protocol B, where the auxiliary rotation occurs around the y-axis, the fringes are narrowed by a factor of $N$ for one parity, and by a factor of $\sqrt{N}$ for the other parity. Both protocols can be modified in a manner that reverse the behavior of the signals for the two parities. We describe an experimental approach where atoms are first caught in a magneto-optic trap, followed by polarization gradient cooling and evaporative cooling, then pushed out in a sequence, and passed through seven interaction zones: three for the conventional CRAIN process, two for auxiliary rotations, and two for one axis twist squeezing, produced via interaction with a detuned probe in a cavity. Over repeated measurements under which the probability of being even or odd is equal, the averaged sensitivity is smaller than the HL by a factor of $\sim$ $\sqrt{2}$ for both versions of the protocol. We describe potential limitations of the proposed approach due to experimental constraints imposed by the current state of the art, for both collective state detection and one-axis-twist squeezing. We show, in Appendix A, the physical interpretation of why the phase magnification in the SCAIN, when configured as a gyroscope, is due to an enhancment of the Compton frequency by a factor of $N$.  On the other hand, we show, also in Appendix A, that when the SC interferometer is configured as an accelerometer, the phase magnification is due to an enhancement of the effective two-photon wave vector by a factor of $N$, leading to the same degree of enhancement in sensitivity.  We also show that such a mesoscopic single object can be used to increase the effective base frequency of an atomic clock by a factor of $N$, with a sensitivity that is equivalent to the HL, within a factor of $\sim$ $\sqrt{2}$. The scheme for this Schr\"odinger Cat Atomic Clock (SCAC) is described in Appendix B.

\acknowledgments{This work has been supported by the NSF grants number DGE-$0801685$ and DMR-$1121262$, and AFOSR grant number FA$9550$-$09$-$01$-$0652$.}

\appendix

\section{Different Physical Interpretations for Phase Magnification in SCAIN for Different Modes:  Enhancement of Compton Frequency for Gyroscopy and Enhancement of Effective Two-Photon Wave Vector for Accelerometry}
For a gyrosocope based on a planar Mach-Zehnder interferometer, a rotation normal to its plane causes a phase shift $\Delta\phi$ that is proportional to the rotation rate $\Omega$, due to the Sagnac effect~\cite{Malykin,Sagnac}. To derive the phase shift, one can compute the Sagnac path difference of the two arms, given by $\Delta L_{S}=2A\Omega/v_p$, where $v_p$ is the phase velocity of the waves propagating along the two arms, and $A$ is the area of the interferometer. The phase shift is then given by multiplying this path difference by the wave vector. Alternatively, one can compute the Sagnac time delay between the two paths, which is found to be $\Delta T_{S}=2A\Omega/c^2$, where $c$ is the vacuum speed of light. It should be noted that this delay is a geometric property of the interferometer loop~\cite{Comment2}, and the parameter $c$ appears in this expression due to the use of the relativistic formula for addition of velocities, having nothing to do with the velocity of the waves propagating along the two arms~\cite{Malykin}. The phase shift is then given by multiplying this time delay by the angular frequency.  For an optical gyroscope, the wave vector and the angular frequency are simply related by the speed of light, and it is easy to see the equivalence between these two methods. However, for a matter-wave gyroscope, the relationship between these two approaches is less obvious. 

To elucidate the equivalence of these two approaches for matter waves, note first that in this case the angular frequency is given by the Compton frequency $w_c$, defined as $E/\hbar$, where $E$ is the relativistic energy of the particle, while the wavevector is $\vb{k}_{dB}$, which is $2\pi$ times the inverse of the de Broglie wavelength, and is given by $\vb{p}/\hbar$, where $\vb{p}$ is the relativistic momentum of the particle. These two quantities are related by the Lorentz transformation~\cite{Haslett,Lan,COSAIN}. It is well-known that $E/c$ and $\vb{p}$ form a four-vector; as such, $w_c/c$ and $\vb{k}_{dB}$ also form a four-vector. In the rest frame of the particle, we have $E=mc^2$, $\vb{p}=0$, where $m$ is the rest mass of the particle. In the frame where the particle is moving at velocity $\vb{v}$, using Lorentz transformation, we have $E=\gamma mc^2$, $\vb{p}=\gamma m\vb{v}$, where $\gamma=1/\sqrt{1-(v/c)^2}$. Therefore for a moving particle, we have $w_c=\gamma mc^2/\hbar$ and $\vb{k}_{dB}=\gamma m\vb{v}/\hbar$. It then follows that the phase shift for the two approaches yield the same value: $\Delta\phi=w_c\Delta T_{S}= k_{dB}\Delta L_{S} = 2mA\Omega/\hbar$, where we have assumed $v \ll c$ so that $\gamma \approx 1$. 

To see transparently why the fringes are amplified by a factor of $N$ for the SCAIN, we recall first that the ensemble can always be viewed as a single particle with a mass of $Nm$, even when there is no entanglement, if a description based on collective states is employed. This was illustrated in our earlier paper on the COSAIN~\cite{COSAIN}, for which the experimental configuration is identical to that of a CRAIN, as discussed in Section~\ref{CRAIN_COSAIN}. For the CRAIN as well as the COSAIN, the sum of the quantum states of $N$ atoms can be expressed, equivalently, as the sum of $N$ collective states, each of which has a mass of $Nm$. The trajectory of each of these collective states during the traversal through the interfermeter depends on the momentum imparted to it, which in turn depends on the fraction of atoms that are in the spin-up state. As such, there are many closed-loops, each with a different effective area. Thus, the fringe pattern for each of these loops has a different width. The final quantum state represents interference between all the collective states. If the population of one of the collective states (e.g., the one where all atoms are in the spin-down state) is detected, as in the case of the COSAIN, then the resulting fringes become akin to that of a Fabry Perot interferometer, and the central fringe is narrowed by a factor of $\sqrt{N}$ compared to the width of the fringes observed in a CRAIN. In the case of the SCAIN, there is only one closed loop, because the quantum state is a superposition of only two collective states. The area of this loop is the same as that for each atom in a CRAIN. However, the mass of each of these two collective states is $Nm$. As such, the Compton frequency for each of these two collective states is amplified by a factor of $N$. Alternatively, the de Broglie wavelength for each of these two collective states is reduced by a factor of $N$.  For either view, it then follows immediately that the phase is magnified by a factor of $N$. The discussion in the preceding paragraph shows that these two views are equivalent, since the spatial phase variation due to the de Broglie wavelength is merely a Lorentz transformation induced manifestation of the temporal phase variation due to the Compton frequency in the rest frame of the particle. However, the interpretation based on the de Broglie wavelength is somewhat misleading, since the actual phase shift does not depend on the velocity --- and, therefore the de Broglie wavelength --- of the particle. The interpretation based on the Compton frequency makes it manifestly obvious that the phase shift has no dependence on the velocity of the particle. 

Next, we note that when a CRAIN is used for measuring acceleration rather than rotation, the phase shift is given by $\Delta \phi=k_{eff}gT^2$, where $k_{eff}=k_{1}+k_{2}$ is the effective two-photon wave vector, given by the sum of the wave vectors for the two legs of the $\Lambda$ transition, and $T$ is the interaction time. This result can be understood by noting that in this case what is measured are the phases of the laser fields. In the rotating waves picture, which is akin to the use of atoms dressed with photons, the spin-down state is dressed by the photon with wave vector of $k_{1}$, while the spin-up state is dressed with a counter-propagating photon with wave vector of $k_{2}$; as such, the phase difference between the dressed spin-up state and the dressed spin-down state is the difference of the phase variations of the two counter-propagating photons, at the spatial rate of $k_{eff}$. It can be shown that, for the SCAIN, the phase shift for the interferometer is amplified by a factor of N: $\Delta \phi=N k_{eff}gT^2$. This is because the collective state $E_{0}$ is dressed by N photons, each with a wave vector of $k_{1}$, while the collective state $E_{N}$ is dressed by N photons, each with a wave vector of $k_{2}$; as such, the phase difference between the dressed collective state $E_{N}$ and the dressed collective state $E_{0}$ is the difference of the phase variations of N pairs of counter-propagating photons, at the spatial rate of $Nk_{eff}$. For both CRAIN and SCAIN, when used for accelerometry, the interferometer phase shift has no dependence on the mass of the atoms; as such, the Compton frequency plays no role in either case.

In this context, it is relevant to note the recent controversy surrounding a paper~\cite{Muller} in which the measurement using a CRAIN, operating as an accelerometer, was re-interpreted as a measurement of gravitational redshift of a clock operating at the Compton frequency of a single atom. While the authors of references~\cite{Wolf,Wolf1,Schleich,Schleich1} dispute this re-interpretation, the authors of the original paper stand by their claim~\cite{Muller1}. If the authors of the original paper are correct, then it follows that the Compton frequency can be used to interpret the signal for a CRAIN/SCAIN even when measuing acceleration. On the other hand, if the objecting authors are correct, then we conclude that the use of the Compton frequency is irrelevant and unnecessary for determining the signal for the CRAIN/SCAIN when measuring acceleration; this is in keeping with the arguments we presented in the preceding paragraph. However, either of these conclusions is irrelevant when considering the use of the CRAIN/SCAIN for measuring rotation; in that case, it is clear that the use of Compton frequency is valid, based on the arguments we presented above. 

Finally, it should be noted that, for a conventional Raman Ramsey Atomic Clock (RRAC), as summarized in Appendix B (where we describe the RRAC as a background for describing the Schr\"odinger cat version thereof), the phase shift is given by $\Delta \phi=2\pi f T_{D}$, where $f$ is the clock-detuning (in Hz), and $T_{D}$ is the time separatio between the two Ramsey zones. This can be viewed as resulting from the fact that (in the rotating waves picture, which is akin to the use of atoms dressed with photons) the spin-down state is dressed with a photon at frequency $f_{1}$, corresponding to one leg of the $\Lambda$ transition, while the spin-up state is dressed with a photon at frequency $f_{2}$, corresponding to the other leg of the $\Lambda$ transition. As such, the clock frequency is defined by the difference between the frequencies of the two photons: $f_{clk}=f_{1}-f_{2}$. At the same time, the energy difference between the bare atom in spin-up and spin-down states is $h f_{atm}$, so that the net energy difference between the dressed spin-up state and the dressed spin-down state is given by $h f=h (f_{clk}-f_{atm})$, which in turn implies a clock detuning of $f$. As shown in detail in Appendix B, for the Schr\"odinger Cat Atomic Clock (SCAC), the phase shift is amplified by the factor N: $\Delta \phi=2\pi N f T_{D}$. This is because the collective state $E_{0}$ is dressed by N photons, each with a frequency $f_{1}$, while the collective state $E_{N}$ is dressed by N photons, each with a frequency $f_{2}$. As such, the energy difference between the dressed $E_{N}$ state and the dressed $E_{0}$ is $N h f=N h (f_{clk}-f_{atm})$, which implies a clock detuning of $N f$. For both RRAC and SCAC, the phase shift does not depend on the atomic mass; as such, the Compton frequency plays no role for either version of the clock.

\section{Schr\"odinger Cat Atomic Clock}
In this appendix, we present the results obtained by applying the proposed protocols to atomic clocks. As mentioned in the main body of the paper, the combination of one axis twist (OAT) spin squeezing, followed by a rotation, inversion of rotation and unsqueezing, along with collective state detection can also be used to realize a parity-independent, mesoscopic Schr\"odinger Cat Atomic Clock (SCAC) with Heisenberg Limited sensitivity, within a factor of $\sqrt{2}$. In order to describe how the SCAC works, we consider first a configuration where the ground states $\ket{\downarrow}$ and $\ket{\uparrow}$ of a three-level atom interact with an excited state $\ket{e}$ via two copropagating laser beams. One of the beams, detuned from resonance by $\delta_1$ and with Rabi frequency $\Omega_1$, couples $\ket{\downarrow}$ to $\ket{e}$. The other beam, with Rabi frequency $\Omega_2$ and detuning $\delta_2$, couples $\ket{e}$ to $\ket{\uparrow}$. For $\delta \gg \Omega_1, \Omega_2, \Gamma$, where $\delta=(\delta_1 + \delta_2)/2$, and $\Gamma$ is the decay rate of $\ket{e}$, the interaction can be described as an effective two level system excited by an effective traveling wave with a Rabi frequency $\Omega=\Omega_1\Omega_2/2\delta$, and detuning $\Delta=\delta_1 - \delta_2$. It should be noted that this is formally equivalent to a conventional microwave atomic clock that couples $\ket{\downarrow}$ to $\ket{\uparrow}$. However, since a Raman transition is needed for the detection of collective states, we choose to describe it here as a Raman clock. In practice, all results presented here would remain valid for a conventional microwave excitation, which is preferable because a Raman clock may suffer from fluctuations in light shifts.

In a conventional Raman Ramsey atomic clock (RRAC), an ensemble of $N$ effective two-level atoms is first prepared in the CSS, denoted as $\ket{-\vu{z}}\equiv\ket{E_0}=\prod_{i=1}^N\ket{\downarrow_i}$. The initial $\pi/2$-pulse rotates the CSS about the $\vu{x}$-axis and brings it to the $\vu{y}$-axis, producing the state $e^{-i(\pi/2) J_x}\ket{-\vu{z}}=\ket{\vu{y}} =\prod_{i=1}^N(\ket{\downarrow_i}-i\ket{\uparrow_i})/\sqrt{2}$. The collective spin is then left to evolve without any interaction for time $T_D$, during which each constituent spin acquires a phase $\phi=2\pi fT_D$, where $f=\Delta/2\pi$ is the (two-photon) detuning of the clock in Hertz. This is equivalent to a rotation by $\phi$ about the $\vu{z}$-axis. At this point, a second $\pi/2$-pulse is applied, which establishes the final state, $\ket{\psi}=\prod_{i=1}^N((1-e^{i\phi})\ket{\downarrow}-i(1+e^{i\phi})\ket{\uparrow})/2$. The aim of the RRAC is to measure $\phi$, and therefore, $f$ as precisely as possible.

In an ideal RRAC, $\phi$ is measured by mapping it onto the operator representing the difference in spin-up and spindown populations: $\hat{J}_z$. The signal, which is a measure of the population of $\ket{\uparrow}$ is, therefore, $S_{RRAC}=J + \langle \hat{J}_z \rangle=N\cos^{2}(\phi/2)$. The associated quantum projection noise is $\Delta S_{RRAC}=\Delta \hat{J}_z=\sqrt{N/4}\sin(\phi)$. The stability of the measurement of $f$ is an indicator of the performance of an atomic clock. The stability of the clock is attributed to the quantum fluctuation in frequency (QFF), analogous to the QFR described in the main body of this paper. The QFF can be written as
\begin{align}
QFF=\Delta f=\left|\frac{\Delta J_z}{\partial \langle J_z \rangle/\partial f}\right| \nonumber\\
=\left(2\pi T_D\sqrt{N}\right)^{-1}\equiv\gamma/\sqrt{N}.
\end{align}
where $\gamma$ is the width of the RRAC fringes.

As is the case for a COSAIN, the COSAC differs from a conventional RRAC in that the measurement of the signal is done on a collective state of the ensemble, instead of single atom measurements~\cite{COSAC}. In the picture based on collective states (which is equivalent to the picture based on individual atoms), the first $\pi/2$-pulse couples the initial state $\ket{E_0}$ to $\ket{E_1}$, which in turn is coupled to $\ket{E_2}$, and so on, effectively causing the ensemble to split into $N+1$ states. During the dark zone, the $n$-th collective state $\ket{E_n}$ picks up a phase $e^{-in\phi}$. When the ensemble interacts with the last $\pi/2$-pulse, each of the collective states interfere with the rest of the states. The COSAC can, thus, be viewed as the aggregation of interference patterns due to ${N+1}\choose 2$ RRAC's working simultaneously. The mathematical derivation of this mechanism is discussed in detail in Ref~\cite{COSAC}. The narrowest constituent signal fringes are derived from interferences between states with the largest difference in phase, i.e. $\ket{E_0}$ and $\ket{E_N}$. The width of this fringe is $\gamma/N$. The widths of the rest of the signal components range from $\gamma$ to $\gamma/(N-1)$. The signal, which is the measure of the population of $\ket{E_N}$, is the result of the weighted sum of all the pairwise interferences with this state. This is detected by projecting the final state of the ensemble, $\ket{\psi}$ on $\ket{E_N}$. Thus, $S_{COSAC}=\langle\hat{Q}\rangle=\cos^{2N}(\phi/2)$, where $\hat{Q} \equiv\ket{E_N}\bra{E_N}$. The quantum projection noise is the standard deviation of $\hat{Q}$, given by $\Delta S_{COSAC}=\cos^{N}(\phi/2)\sqrt{1-\cos^{2N}(\phi/2)}$. The QFF of the COSAC is thus,
\begin{align}
\Delta f\bigr|_{COSAC} & = \left|\Delta \hat{Q}/\partial_{f} \langle \hat{Q}\rangle \right|\nonumber\\
& = (\Delta f\bigr|_{CRAIN}/\sqrt{N})|\sqrt{\sec^{4J}(\phi/2)-1}/\tan(\phi/2)|
\label{Eq: COSAC QFF}
\end{align}
Therefore, for $f \rightarrow 0$, the frequency sensitivity of the COSAC is the same as that of an RRAC, assuming that all the other factors remain the same. 

The SCAC is based on the same process of squeezing followed by a rotation and then another rotation and unsqueezing as that employed for the SCAIN. The CSS after the first $\pi/2$-pulse is squeezed via the OAT spin squeezing Hamiltonian, $H_{OAT} = \hbar \chi J_z^2$, yielding the SSS of the ensemble $\ket{\psi_e}=e^{-i \mu J_z^2}\ket{\vu{y}}$, where $\mu=\chi\tau$ is the squeezing parameter, and $\tau$ is the duration of the squeezing interaction. This SSS must be rotated by an angle $\nu$ about an appropriate axis, the choice of which depends on the degree of squeezing, and follows the same rules as described in the main body of the paper.

Similar to the SCAIN, the SCAC can be operated under two different protocols, which are essentially identical, except for the choice of the axis around which we apply a rotation that maximizes the degree of observed squeezing, and the amount of the rotation. In one case (Protocol A), the rotation is around the $\vu{x}$ axis, and the amount of rotation is always $\pi/2$. In the other case (Protocol B), the rotation is around the $\vu{y}$ axis, and the amount of rotation depends on the degree of squeezing. We first consider Protocol A, focusing initially on the special case where $\mu=\pi/2$, with the case of an arbitrary value of $\mu$ to be discussed later. For even $N$, $H_{OAT}$ transforms $\ket{\vu{y}}$ to $\ket{\psi_e} = (\ket{\vu{y}} - \eta \ket{-\vu{y}})/\sqrt{2}$, where $\eta=i(-1)^{N/2}$, representing a phase factor with unity amplitude. As we noted in the main body of this paper, this phase factor depends on the Super Even Parity (SEP); however, the shapes of the fringes, as well as the values of QFF, are not expected to depend on the value of the SEP, as we have verified explicitly. Rotating $\ket{\psi_e}$ by $\nu=\pi/2$ about the $\vu{x}$ axis yields the Schr\"odinger cat state $\ket{\psi_{SC}}=(\ket{E_0} + \eta \ket{E_N})/\sqrt{2}$. At the end of the dark zone, the state of the ensemble is $ (e^{iN\phi/2}\eta \ket{E_N} + e^{-iN\phi/2}\ket{E_0})/\sqrt{2}$. We now apply a rotation of $\nu=\pi/2$ about the $\vu{x}$ axis (Ideally inversion of the rotation would require the application of rotation of $\nu=-\pi/2$. However, we have found~\cite{Shahriar2} that changing the sign of this rotation simply inverts the final fringes. This is also true for the SCAIN protocol. It should also be noted that experimentally, $\nu=-\pi/2$ actually corresponds to $\nu=3\pi/2$, which requires a longer duration or more power. Therefore, for both the SCAIN and the SCAC, we choose to use a corrective rotation of $\pi/2$ rather than $-\pi/2$), followed by the untwisting Hamiltonian, $-H_{OAT}$. Finally, the last $\pi/2$ pulse is applied to catalyze interference between the resulting states. The signal arising from this interference depends on $\phi$ as $S_{SCAC} =\langle\hat{Q}\rangle= \sin^2(N\phi/2)$.

When $N$ is odd, initial squeezing produces $\ket{\psi_e}=(\ket{\vu{x}} +\rho \ket{-\vu{x}})/\sqrt{2}$, where $\rho=i(-1)^{(N+1)/2}$, representing a phase factor with unity amplitude. As noted in the main body of the paper, this phase factor depends on the Super Odd Parity (SOP); however, the shapes of the fringes, as well as the values of QFF, are not expected to depend on the value of the SOP, as we have verified explicitly. For $\phi=0$, the sequence $e^{-i\nu J_x}e^{-i\phi J_z}e^{-i\nu J_x}$ causes a $\pi$ phase-shift in each of the components of this state. Application of the unsqueezing Hamiltonian, $-H_{OAT}$ then moves the system to $\ket{-\vu{y}}$, and the final $\pi/2$ pulse places the system in the $\ket{-\vu{z}}$ state, which is the same as the collective state $\ket{E_0}$. Since we detect the collective state $\ket{E_N}$, the whole sequence thus generates a null signal. Again, just as in the case of the SCAIN, the same conclusion holds for an arbitrary value of $\phi$, for reasons that are not manifestly obvious due to the complexity of the states, but can be verified via simulation. Over repeated measurements, the probability of $N$ being even or odd is equal. Thus, for $M$ trials, the average signal of the SCAC in this regime is $S_{SCAC}=M\sin^2(N\phi/2)/2$. The associated quantum projection noise is $\Delta S_{SCAC}=\sqrt{M/2}\sin(N\phi)$. The QFF is thus, $\Delta f=1/\sqrt{2M}\pi NT_D$, which is a factor of $\sqrt{2}$ below the HL.

\begin{figure}[h]
\includegraphics[scale=0.53]{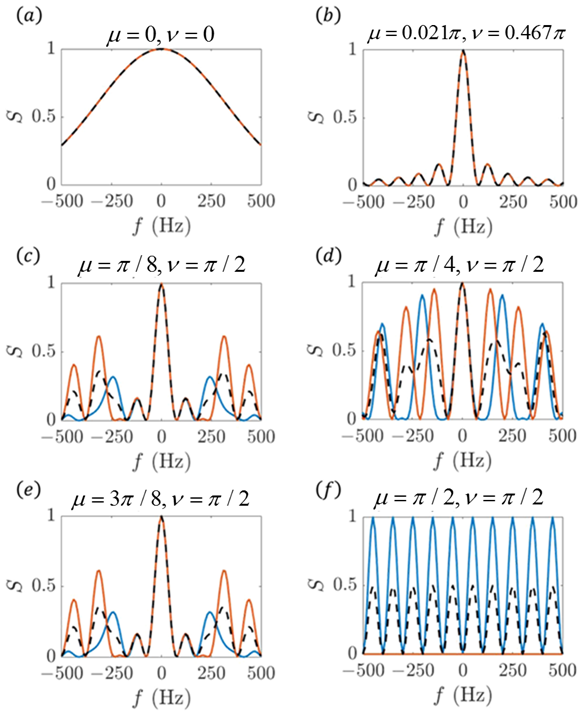}
\caption{Signal fringes for various values of $\mu$, $T_D=\SI{50}{\micro\second}$. $N=200$ is indicated by blue lines, $N=201$ by red lines. The broken black lines indicate the average signal. Figures (a)-(e) employ Protocol B, while figure (f) employs Protocol A. The time interval between the $\pi/2$ pulses is $\SI{50}{\micro\second}$, so that the peak-to-peak width of a conventional clock fringe would be $\SI{20}{\kilo\hertz}$. The peak-to-peak width of the blue fringes in figure (f) is seen to be $\SI{100}{\kilo\hertz}$, corresponding to a factor of $N$ reduction for Protocol A.}
\label{app2-fig:1}
\end{figure}

Next, we consider Protocol B, in which the rotation is always around the $\vu{y}$ axis while the rotation angle $\nu$ is chosen so as to maximize (right after the squeezing interaction) the fluctuations along the $\vu{z}$ axis. For a given value of $N$, $\nu$ increases with $\mu$, reaching a maximum value of $\pi/2$ at $\mu=\mu_0$ ($\mu_0=0.095\pi$ for $N=200$). Once the SSS is optimally aligned, the dark zone follows. We now apply another rotation $-\nu$ about the $\vu{y}$ axis (note that this rotation is a reversal of the original rotation, unlike the case for Protocol B in SCAIN), then apply $-H_{OAT}$. Finally, the last $\pi/2$ pulse is applied to establish the final state. This signal fringes as a function of $\phi$ under Protocol B are illustrated in Fig.~\ref{app2-fig:1}~(a)-(e), for various values of $\mu$. The results for even values of $N$ ($N=200$) are indicated by the blue lines, and those for the odd values of $N$ ($N=201$) are indicated by the orange lines. The broken black lines indicate the average signal. Until the value of $\mu$ gets close to $\pi/2$, the central fringe as a function of frequency is essentially identical for both odd and even values of $N$. Thus, for $M$ trials, the average signal is independent of the parity of $N$ for the central fringe, which is the only one relevant for metrological applications. For different values of $\mu$, the non-central fringes, averaged over the odd and even cases, have different shapes, heights and widths. However, the central fringe always has full visibility. Its width first decreases sharply with increasing values of $\mu$, and then saturates at $\mu=\mu_0$. Consequently, the fluctuations in frequency drops significantly, attaining the minimum value $\Delta f|_{SCAC}=e^{1/3}/\sqrt{M}2\pi N T_D$, at $\mu=\mu_0$.

For the limiting case of $\mu=\pi/2$, Protocol B produces very different results for odd and even values of $N$, as shown in Fig.~\ref{app2-fig:1}~(e). Specifically, for odd values of $N$, this protocol produces uniform fringes, each with a width that is factor of $N$ narrower than what is observed in a conventional RRAC, thus yielding HL sensitivity. In this case, the ideal Schr\"odinger Cat state is realized, in a manner analogous to what we described above for Protocol A (with $\mu=\pi/2$). For odd values of $N$, this protocol also produces uniform fringes, but each with a width that is the same as that observed for COSAC (which is a factor of $\sqrt{N}$ narrower than what is observed in an RRAC), thus yielding SQL sensitivity. The average of these two signals, for many repeated measurements, would produce a sensitivity that, for large $N$, is lower than the HL by a factor of $\sqrt{2}$~\cite{Shahriar2}. In Fig.~\ref{app2-fig:1}~(f), we show the corresponding fringes produced using Protocol A, for the special case of $\mu=\pi/2$. As described earlier, in this case, we get a purely sinusoidal fringe pattern for even values of $N$, and a null signal for odd values of $N$. The averaged signal, therefore, is also purely sinusoidal. The width of these fringes is a factor of $N$ narrower than what is observed in a conventional RRAC.

\begin{figure}[h]
\includegraphics[scale=0.40]{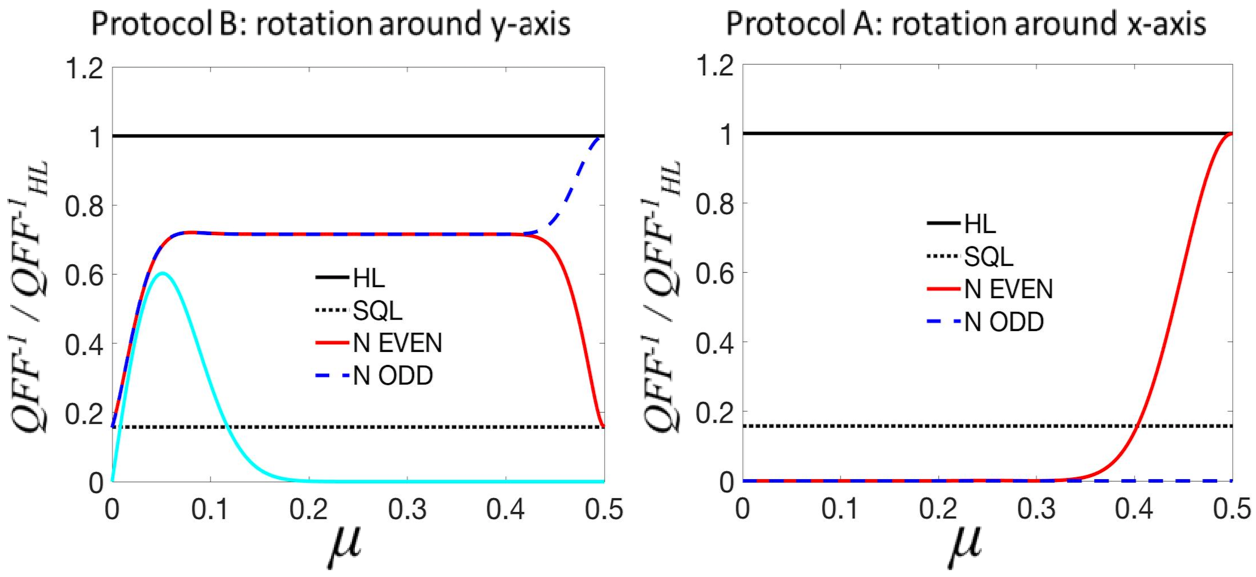}
\caption{QFF$^{-1}$ of SCAC vs the squeezing parameter, $\mu$, normalized by the same for the HL for $N=100$. Horizontal lines indicate the HL (black solid), and the SQL (black dashed). The dashed blue lines corresponds to odd value of N ($N=101$) and the red lines corresponds to even value of N ($N=100$). The left(right) panel shows the results for Protocol B(A). The cyan line in the left panel shows the corresponding result for the squeezing-unsqueezing protocol proposed in Ref.~\cite{Antisqz} and demonstrated subsequently in Ref.~\cite{Antisqzexpt}.}
\label{app2-fig:2}
\end{figure}

In Fig.~\ref{app2-fig:2}, we summarize the results for both protocols, for squeezing parameters ranging from $\mu=0$ to $\mu=\pi/2$. The behavior is essentially identical to that shown in Fig.~\ref{fig:5} in the main body of the paper for the SCAIN. Here, we show the inverse of the QFF, normalized to the same for the HL for $N=100$, as a function of $\mu$. Horizontal lines indicate the HL (black solid), and the SQL (black dashed). The blue lines corresponds to odd value of N ($N=101$) and the red lines corresponds to even value of N ($N=100$). The left panel shows the result of using Protocol B. The value of {QFF$^{-1}$ increases monotonically, reaching a peak value at $\mu=\mu_0$, and then remains flat until getting close to $\mu=\pi/2$, with virtually no difference between the odd and even values of $N$. Near $\mu=\pi/2$, the value of {QFF$^{-1}$ begins to diverge, reaching the HL (SQL) for odd (even) values of $N$ at $\mu=\pi/2$. The cyan line in the left panel shows, for comparison, the corresponding behavior of the squeezing-unsqueezing (SU) protocol recently proposed in Ref.~\cite{Antisqz} and demonstrated subsequently in Ref.~\cite{Antisqzexpt}. This protocol also produces a sensitivity close to the HL, but only for a particular value of $\mu$, and then drops off rapidly for both decreasing and increasing values of $\mu$. In contrast, the Protocol B proposed here reaches a sensitivity that is slightly higher than that attainable for the SU protocol, and is highly insensitive to the precise value of $\mu$ after reaching the plateau, as shown in the left panel of Fig.~\ref{app2-fig:2}. The right panel shows the result of using Protocol A. At $\mu=\pi/2$, {QFF$^{-1}$ is at the HL for even values of $N$, and vanishes for even values of $N$. For $\mu < \pi/2$, the amplitude of the signal for even values of $N$ decreases rapidly, with corresponding decrease in the value of {QFF$^{-1}$. Just as in the case of the SCAIN, the vanishing value of {QFF$^{-1}$ is due simply to the vanishing of the signal itself.

\section{List of Abbreviations}
In this appendix, we list all the abbreviations used in this paper in alphabetical order.

AI: Atomic Interferometer; AOM: Acousto-Optic Modulator; ARA: Auxiliary Rotation Axis; BEC: Bose-Einstein Condensation; CD: Conventional Detection;CD-SCAIN: Conventional Detection - Schr\"odinger Cat Atomic Interferometer; COM: Center of Mass; COSAC: Collective State Atomic Clock; COSAIN: Collective State Atomic Interferometer; CRAIN: Conventional Raman Atomic Interferometer; CSD: Collective State Detection; CSD-SCAIN: Collective State Detection - Schr\"odinger Cat Atomic Interferometer; CSS: Coherent Spin State; DCS: Dicke Collective State; EN: Excess Noise; ESP: Echo Squeezing Protocol; FORT: Far-Off Resonant Trap; FS: Frequency Synthesizer; HL: Heisenberg Limit; LPF: Low Pass Filter; OAT: One Axis Twist; QND: Quantum Non-Demolition; QFF: Quantum Fluctuation in Frequency; QFR: Quantum Fluctuation in Rotation-rate; QPD: Quasi Probability Distribution; QPF: Quantum Phase Fluctuation; QPN: Quantum Projection Noise; PF: Phase Fluctuation; PGS: Phase Gradient of the Signal; RRAC: Raman Ramsey Atomic Clock; SC: Schr\"odinger Cat; SCAC: Schr\"odinger Cat Atomic Clock; SCAIN: Schr\"odinger Cat Atomic Interferometer; SDS: Standard Deviation of the Signal; SEP: Super Even Parity; SNR: Signal to Noise Ratio; SQL: Standard Quantum Limit; SOP: Super Odd Parity; SSS: Squeezed Spin State; SU: Squeezing-Unsqueezing; TACT: Two-Axis-Counter-Twist; XDCS: X-directed Dicke Collective State; YDCS: Y-directed Dicke Collective State; ZDCS: Z-directed Dicke Collective State


\bibliographystyle{apsrev}

\begin{thebibliography}{100}
\bibitem{RefComment} The signal corresponding to an atomic interferometer can also be derived using de Broglie wavelength. However, it can be shown that this approach simply corresponds to interference at the Compton frequency in a moving frame, as shown in Appendix A and Refs.~\cite{Lan, COSAIN}.
\bibitem{CommentVariance} In the literature, such as referencese ~\cite{Leroux1,Monika,Thompson,Hosten}, it is customary to report the suppression in variance. However, the actual factor of improvement in measurement sensitivity is proportional to the square-root of the variance, if the signal remains unchanged. For protocols such as the one proposed here, the improvement is due to a combination of reduction in variance, reduction in signal, and phase magnification of the signal. As such, we define the improvement in terms of the Quantum Fluctuation in Rotation-rate (QFR) for the case of gyroscopes, and Quantum Fluctuation in Frequency (QFF) in the case of clocks. For the protocol used here, when averaged over odd and even values of atoms, assumed to occur with equal probabilities, the improvement in QFR or QFF is by a factor of $\sqrt{N/2}$. If the improvement were due to an equivalent reduction in variance only, the reduction would be by a factor of ${N/2}$. Thus, for $N=5 \times 10^{5}$, the equivalent reduction in variance for our protocol would be 54 dB.
\bibitem{Comment2} The Sagnac time delay between the two paths is generally computed by determining the velocities of the waves in the two arms, as seen in the reference rest frame, by using relativistic expressions for additions of velocities. Not surprisingly, it can also be derived using Lorentz transformation directly. The time delay for the counter-clock-wise (CCW) path is $\Delta T_{0}=\Delta T_{S}/2=A\Omega/c^2$, while that for the clock-wise (CW) path is $-\Delta T_{0}$. Consider a differential section of length $\delta L_{0}$ along the CCW path, for example. The difference between the end points of this section, at a given time in the rest frame for this section, represents a proper length. The time difference between these two events, in the reference rest frame, is given by $v \delta L_{0}/c^2$, where v is the speed of this section with respect to the reference rest frame: for a circular path, $v=\Omega R$, where R is the radius of the circle. This time difference represents the time delay for this segment. The total time delay for the CCW path is just the sum of these delays, since each differential segment has the same speed. Thus, the total delay for the CCW path, of length $L_{0}=\pi R$ is $\Delta T_{0}=v L_{0}/c^2=\Omega R \pi R/c^2=A\Omega/c^2$. For the CW path, the motion is in the opposite direction, thus reversing the sign of the delay. It is easy to show that the same expression also holds for the trapezoidal shape of the atomic interferometer. Thus, the Sagnac time delay is a simple manifestation of the non-simultaneity of events in one frame that are simultaneous in another frame, and is purely geometric in nature, with no dependence on the velocity of the waves propagating along the interferometer.
\end{thebibliography}

\end{document}